\newcommand{\beq}{\begin{equation}}
\newcommand{\eeq}{\end{equation}}
\newcommand{\beqa}{\begin{eqnarray}}
\newcommand{\eeqa}{\end{eqnarray}}

\documentclass{ws-p9-75x6-50}

\begin{document}

\title{Pairing in Inhomogeneous Superconductors }
\author{J. Eroles$^{1,2}$, G. Ortiz$^{1}$, A. V. Balatsky$^{1}$ and A.
R. Bishop$^{1}$}
\address{$^1$Theoretical Division, 
Los Alamos National Laboratory, Los Alamos, NM 87545, USA.\\
$^2$Centro At\'{o}mico Bariloche and Instituto Balseiro, 
S. C. de Bariloche, Argentina.\\}
\maketitle

\abstracts{
Starting from a $t$-$J$ model, we  introduce inhomogeneous terms
 to mimic stripes. We find that if the inhomogeneous terms break
the $SU(2)$ spin symmetry the binding between holes is tremendously
enhanced in the thermodynamic limit. In any other model (including
homogeneous models) the binding in the thermodynamic limit is small or
neglible. By including these inhomogeneous terms we can reproduce experimental
neutron scattering data.  We also discuss the connection of the
resulting inhomogeneity-induced superconductivity to recent
experimental evidence for a linear relation between  magnetic
incommensurability and the superconducting transition temperature, as a
function of doping. }

\section{Introduction}

Understanding the high-temperature superconducting cuprates remains a major
goal in quantum many-body physics. Every conventional approach has failed to
adequately explain their normal and superconducting phases. This fact, together
with some new experimental data, may indeed point to the need for a new
conceptual frame. In this work we propose a scenario  based
on three basic assumptions: {\bf 1}- The superconducting state is
inhomogeneous. {\bf 2}- At the inhomogeneities (stripe segments), the
spin-rotational symmetry is broken,  providing a background for
the charge carriers to form bound pairs. {\bf 3}- These pairs Josephson-tunnel
between stripes. In this scenario, there are (at least) two different energy
scales: A lower one related to the phase coherence of the superconducting state
(and therefore to $T_c$) and another related to the pairing of holes. In this
regard, there are some similarities to granular superconductors.

Recent neutron, X-ray, Raman and phonon-measurements
\cite{phonons}, strongly suggest that at low temperatures and moderate doping the
system is spatially inhomogeneous. The simplest realization of these
inhomogeneities are termed ``stripes.''\cite{zannen} In these stripes, charge clusters into
nanoscale one-dimensional (1D) structures while the rest of the material
displays strong antiferromagnetic correlations. There is no phase separation:
the stripes are spatially separated. Note that this scenario is quite different
from the BCS one, where the formation of a
homogeneous superconducting state is described with a homogeneous superfluid
density. 

Remarkably, there is a subtle interplay between magnetism and superconductivity.
In the undoped case, neutron scattering experiments
show a peak at $\bf{k}=\bf{Q}=(\pi,\pi)$. At finite hole doping, this peak splits
into four (at a distance $(\pm \delta,\pm \delta)$ from $\bf{Q}$), indicating the
formation of antiferromagnetic domains. $\delta$ increases with  hole
concentration, suggesting that the stripes come closer. (It is believed that the
concentration of holes in the stripe is nearly constant and equal to 1/2.) In the
insulating state the position of these four peaks is such that the stripes run
along a diagonal. Upon increasing the hole concentration further, these peaks
rotate by  $\pi/2$ near the superconducting transition\cite{birgeno1D}. This is
evidence that the two features (superconductivity and stripes) are
inter-related. 
There is also experimental evidence showing that
the stripes are 1D objects\cite{birgeno1D}. The relation between spin
incommensuration and charge ordering has been experimentally shown in
Ref.[1] where, for doping $x=\frac{1}{8}$, X-rays
diffraction displays the same four peaks (with incommensuration
$2\delta$). On the other hand, $T_c$ and $\delta$  seem to be {\it
linearly} related in LSCO and YBCO\cite{Yamada}: $k_B$ $T_c
= \hbar v^* \delta$, where $v^*$ defines a material-dependent velocity
scale. Therefore, the only dependence of $T_c$ on $x$ is through $\delta(x)$. 
There is also experimental evidence
supporting the existence of a spin-gap in these
compounds\cite{spingap}. 

These experimental facts suggest quasi-1D objects, rich in holes, 
separating $\pi$-shifted
antiferromagnetic domains. There has been considerable
theoretical work attempting to prove that a stripe state is the low-energy
state of homogeneous $t$-$J$ or Hubbard models
\cite{White-Scalapino,stripes-Hubb,Manusakis}, i.e., a broken symmetry
state of doped Mott insulators. There is no general
consensus on this issue; it is most probable that the stripe
state is an excited state of those homogeneous models
\cite{Manusakis}. Here, we adopt a different strategy, namely
introducing explicitly inhomogeneous terms in the model (which
break the translational symmetry). We conclude that 
inhomogeneous terms breaking spin-rotational invariance locally are the most
efficient way to produce a bound state of two holes \cite{we}. 

Next we will introduce our microscopic model, discuss
pairing and magnetic properties of the model, and finally will make an
attempt to explain the phase-locked superconducting state using a
phenomenological Josephson-spaghetti model. 

\section{Microscopic Model}

\begin{figure}[t] 
\epsfxsize=25pc 
\centerline{\epsfbox{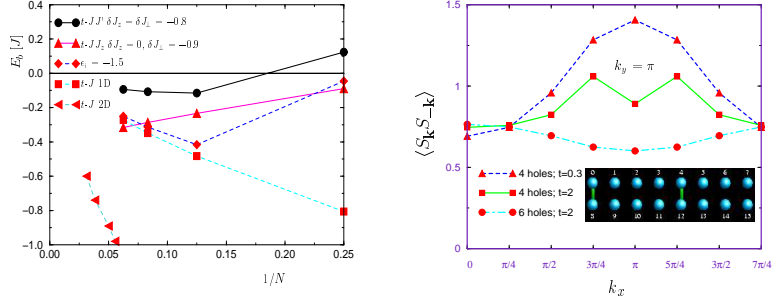}} 
\caption{{\it Left}: Binding energy, $E_b$, for different models in 1D
and 2D. $t$-$JJ'$ is a $t$-$J$ model with two weakened bonds (with
$\delta J_{\perp}=\delta J_z<0$). $t$-$J$$\epsilon_i$ has on site
energy $\epsilon_i < 0 $ on the stripe. $t$-$J$$J_z$ has two easy-axis
spin-rotational symmetry broken bonds every $P=4$ sites. In all cases
$J=t=1$. {\it Right}: Spin structure factor for a $t$-$J$$J_z$ ladder
(8$\times$2) with two $\delta J_{\perp}=-0.9$ bonds in the
Y-direction (see inset). The incommensurability appears only for $t$
larger than a critical value. When 6 holes are added to the system the
double peak disappears and is replaced by a broad one around ${\bf
k}=(0,\pi)$. 
\label{fig1}} 
\end{figure}

Our microscopic scenario\cite{we} starts from a homogeneous $t$-$J$
model as background:
\begin{equation} 
\label{H} 
H_{t\!-\!J} = -t  \sum_{\langle i,j \rangle, \sigma} c^{\dagger}_{i
\sigma} c^{\;}_{j \sigma} + J \sum_{\langle i,j \rangle} ({\bf S}_i
\cdot {\bf S}_j - \frac{1}{4} \bar{n}_i \bar{n}_j )  ,
\end{equation} 

\noindent where $c^{\dagger}_{i\sigma}$ creates a fermion in the space with
double occupancy forbiden, ${\bf S}_i$ is the spin operator and 
$\bar{n}_i=c^{\dagger}_{i\sigma}c_{i\sigma}$
is the number operator. To mimic the stripe segments, we add inhomogeneous
magnetic interactions. These inhomogeneous terms break translational invariance
and spin-rotational $SU(2)$ symmetry locally:

\begin{equation} 
H_{\rm inh} = \sum_{\langle \alpha,\beta \rangle} \delta J_z \
S^z_\alpha S^z_\beta + \frac{\delta J_{\perp}}{2} \left( S^+_\alpha 
S^-_\beta + S^-_\alpha S^+_\beta \right)
\nonumber \ , 
\end{equation} 
with $\delta J_{\perp} \neq \delta J_z$, representing the magnetic 
perturbation of a static local Ising anisotropy, locally lowering spin
symmetry (named $t$-$JJ_z$ model). Only a few links 
$\langle \alpha,\beta \rangle$ (at the stripes) 
have this lowered spin symmetry.

We have studied the binding energy of two holes ($E_b=(E_{2 \, {\rm
holes}}-E_{0 \, {\rm hole}})-2 (E_{1 \, {\rm hole}}-E_{0 \, {\rm hole}})$) for
different 1D and 2D lattices and extrapolated to the thermodynamic limit
(Fig.\ref{fig1} {\it Left}). We conclude that only the $t$-$JJ_z$ model
leads to considerable binding (we have also tried one-band Hubbard models with
many different inhomogeneous terms).  In the bound state, depending upon the
value of $t$, the holes pair in the same or on different stripe segments (in
both cases the binding energy is appreciable). It has been suggested that
homogeneously breaking the spin-rotational symmetry stabilizes the stripe state
\cite{kivelson-Jz}. Here, we argue that doing it inhomogeneously also gives an
excellent hole binding mechanism. One should note that the pair is not
bound to an inhomogeneity: the wave function is 
spread over the whole system, but more concentrated around the stripes.

This model can also explain the magnetic properties outlined above. For
instance, we have calculated the magnetic structure factor in an 8$\times$2
cluster, in which we have placed two stripes by breaking the
spin-symmetry in a $Y$-link every 4 sites (the stripes are
perpendicular to the $X$-axis). Although we
cannot perform a good scaling here, in all cases studied the binding
energy extrapolates to a significant value. The spin structure factor is shown
in Fig.\ref{fig1} {\it Right}. The experimentally observed incommensuration
appears for sufficiently high kinetic energy, $t$. 

Contrary to the homogeneous $t$-$J$ and inhomogeneous $t$-$JJ'$ (where
a link is weakened without breaking the spin-rotational symmetry)
models, the $t$-$JJ_z$ model displays a spin-gap, as seen
experimentally\cite{spingap}. 

For a concentration near optimal doping, the stripe segments are close
enough to losing their identity, suggesting the mechanism for the decrease 
of $T_c$. 

\section{Phenomenological approach to inhomogeneous superfluidity}

\begin{figure}[t]
\epsfxsize=25pc 
\centerline{\epsfbox{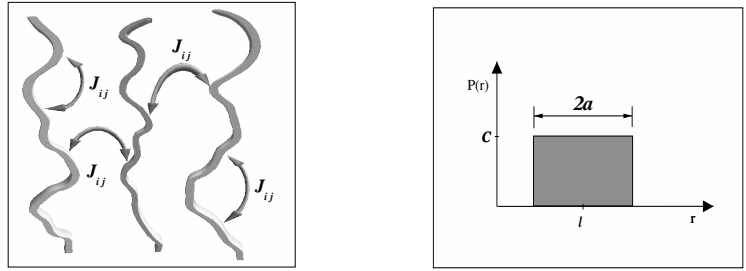}} 
\caption{ Schematic Josephson coupling between an assumed distribution of stripe
segments ({\it Left}) and probability distribution $P(r)$ ({\it Right}). For the
incommensuration $\delta$ to be observed  along crystallographic (1,0) and (0,1)
directions, the stripe-stripe distances must have average $\langle r \rangle
\approx \ell = 1/\delta$. $\langle J \rangle$ is determined by the probability
distribution $P(r)$. Clearly $P(r)$ should be centered near $\ell$, with  some
width ($2a$) from the meandering of stripes and  height $C=[4 \pi \ell
a]^{-1}$ (see text). }
\label{Pr}
\end{figure}

To understand the linear relation between $T_c$ and $\delta(x)$,  based on the
microscopic model phenomenology, we introduce a {\it Josephson Spaghetti
model}\cite{we}. This linear relation could be explained by connecting the
superconducting mechanism to stripe fluctuations\cite{we}. In the following we
consider Josephson tunneling of pairs between stripe segments. The simplest
mean-field model involving {\it only} the phase $\phi(r_i)=\phi_i$ of the order
parameter is
\beqa
{\cal H} = \sum_{ij} J_{ij} \ \exp[i(\phi_i - \phi_j)] \ , \;\;
\mbox{where} \;\; J_{ij} = J(r_{ij}) = t_0/r_{ij}^{\alpha} \ .  \label{ham} 
\eeqa 
The indices $i$ and $j$ stand for coarsed grained regions where the phase is
well defined (around the stripes).  The Josephson coupling is an inter and
intra stripe distance dependant quantity $J(r)$, and the distance $r$ is a
variable with a certain distribution $P(r)$. The mean-field $T_c$ depends upon
the Josephson coupling $\langle J(r) \rangle$. For simplicity we will take
$P(r)$ as the ``box" distribution depicted in Fig. \ref{Pr}. We find

\begin{equation} 
\langle J(r) \rangle =\int d^2 r P(r) J(r) = {2\pi t_0
C\over{2-\alpha}} a_1 \ell^{2-\alpha}  \ , \;\;  \langle r \rangle =
{2\pi C\over{3}} a_2 \ell^3 \ , 
\end{equation} 
with the constants $a_1$ and $a_2$ ${\cal O}(1)$ numbers. Thus, for $\alpha =
1$, we obtain the experimentally observed relation: 
$T_c(x) \simeq \langle J(r) \rangle \propto [\langle r \rangle]^{-1}
=\delta(x)$.


\end{document}